\documentstyle[12pt]{article}

\begin{document}

\begin{center}
{\bf OFF-SHELL NN POTENTIAL AND TRITON BINDING ENERGY}
\footnote{Contributed Paper submitted to the {\it 14th
International Conference
on Few Body Problems in Physics}, Williamsburg, Virginia,
USA, May 26-31, 1994.}
\\
\vspace*{.42cm}
Y. Song
and
R. Machleidt
\\
Department of Physics, University of Idaho,
Moscow, ID 83843, U.S.A.
\\
\vspace*{.42cm}
SUMMARY AND CONCLUSIONS
\end{center}
The ({\it nonlocal\/}) Bonn-B
potential predicts 8.0 MeV binding energy
for the triton (in a charge-dependent 34-channel Faddeev calculation)
which is {\it about 0.4 MeV more than} the predictions by {\it local}
NN potentials.
We pin down origin and size of the nonlocality in the Bonn potential,
in analytic and numeric form.
The nonlocality is due to the use of the correct off-shell Feynman amplitude
of one-boson-exchange avoiding the commonly used on-shell approximations
which yield the local potentials.
We also illustrate how this off-shell behavior leads to more
binding energy.
We emphasize that the increased binding energy is not due to
on-shell differences (differences in the fit of the NN data or
phase shifts). In particular,
the Bonn-B potential reproduces accurately
the $\epsilon_1$ mixing parameter up to 350 MeV as determined in the
recent Nijmegen
multi-energy NN phase-shift analysis.
Adding the relativistic effect from the relativistic
nucleon propagators in the Faddeev equations, brings the Bonn-B
result up to 8.2 MeV triton binding [1]. This leaves a difference
of only 0.3 MeV to experiment, which may possibly be explained by
refinements in the treatment of relativity and the inclusion of
other nonlocalities (e.~g., quark-gluon exchange at short
range).
Thus, it is conceivable that a realistic NN potential which
describes the NN
data up to 300 MeV correctly may explain the triton binding energy
without recourse to 3-N forces; relativity would play a major
role for this result.

\begin{center}
INTRODUCTION
\end{center}

Recently it has been shown that {\it local NN potentials}
lead to a triton binding energy of $7.62\pm 0.01$ MeV [2],
i.~e., they {\it underbind the triton by 0.86 MeV}.
On the other hand, it is well known that the nuclear force
must contain nonlocalities, since any more fundamental
mechanism for creating the nuclear force generates
a nonlocal interaction. Such nonlocalities may
increase binding energy predictions for systems of three or more
nucleons.
An example is the Bonn-B potential [3] which predicts
7.97 MeV binding energy for the triton in a
charge-dependent 34-channel Faddeev
calculation~[4].

Before one can pin down off-shell effects, it is absolutely crucial
to make sure
that the potential under consideration reproduces accurately
the on-shell NN data. Of particular importance is here the
$\epsilon_1$ mixing parameter
which is a measure
for the on-shell tensor force strength. Figure~1 shows
that the the Bonn-B potentials predicts $\epsilon_1$
in accurate agreement with
the most recent Nijmegen multi-energy NN phase-shift
analysis [6].

\begin{center}
OFF-SHELL POTENTIAL AND NN $t$-MATRIX
\end{center}

Predictions for NN observables
may be based upon the on-shell two-nucleon $t$-matrix
derived from a given NN potential $V$.
The off-shell NN $t$-matrix is
input for momentum-space Faddeev calculations
of the three-nucleon system.
The calculation of the $t$-matrix always involves the NN potential
on- and off-(the-energy-)shell. We illustrate this
for the example of the partial-wave $t$-matrix,
$t^{JST}_{L'L}$,
in the $^3S_1$ two-nucleon state,
which is given by
\begin{eqnarray}
t^{110}_{00}({ q'}, { q};E) & = & V^{110}_{00}({ q'}, { q}) -
 \int_0^\infty k^2 dk V^{110}_{00}({ q'}, { k})
\frac{M}{k^2-ME-i\epsilon}
t^{110}_{00}({ k}, { q};E) \nonumber \\
     &    &
- \int_0^\infty k^2 dk V^{110}_{02}({ q'}, { k})
\frac{M}{k^2-ME-i\epsilon}
t^{110}_{20}({ k}, { q};E) \; .
\end{eqnarray}
For free-space NN scattering,
$E=q^2_0/M$ with $M$ the nucleon mass and $q_0$ the c.m.\ on-shell
momentum which is related to the lab.\ energy by $E_{lab}=2q^2_0/M$.
Notice that in the integral terms the potential contributes essentially
off-shell. In general,
the second integral  which involves
$V^{110}_{02}$ (tensor force)
is much larger than the first integral that involves
$V^{110}_{00}$ (central force).
Therefore, we will focus here on the second integral term
and the tensor force.

In Fig.~2, we show the $^3S_1$-$^3D_1$ potential matrix element,
$-V^{110}_{02}(q_0,k)$, for Paris [5] and Bonn [3].
The momentum $q_0$ is held fixed at 153 MeV which corresponds
to a lab.\ energy of 50 MeV.
The abscissa, $k$, is the variable over which the integrations
are performed in the above equation.
It is seen that, particularly for large off-shell momenta,
the Bonn-B potential
is substantially smaller than the Paris potential:
the Bonn-B potential has clearly a weaker {\it off-shell}
tensor force than the Paris
potential.
As a consequence of this, the Paris potential produces
a much larger integral term in Eq.~(1) than the Bonn potential;
and it is this integral term that is subject
to quenching in few- and many-body calculations.
In the three-body Faddeev equations, the $t$-matrix
 is fully off-shell and $E$
is negative; the negative $E$
reduces the magnitude of the integral term.
The larger the term, the larger the quenching.
Since this integral term is attractive, the quenching is a repulsive
effect. Thus, large off-shell potentials, implying a large
integral term, yield less attraction
in three- and many-body problems.
This explains why the Paris potential predicts less triton binding
energy than the Bonn potential.

\begin{center}
THE ORIGIN OF OFF-SHELL DIFFERENCES
\end{center}

The Bonn potential is defined in terms of the full, relativistic Feynman
amplitudes of one-boson-exchange,
which are non-local expressions.
On the other hand,  the Paris potential is (apart from a
${\bf p}^2$-term in the central force)
a local potential.
The local approximation of the pion tensor  potential,
which is used in the Paris potential as well as in any other
local $r$-space potential,
has the familiar form,
\begin{equation}
^\pi \tilde{V}_T(r)=
-\frac{g^2_\pi}{4\pi}
\left(\frac{m_\pi}{2M}\right)^2
\left(1+\frac{3}{m_\pi r}+\frac{3}{(m_\pi r)^2}\right)
\frac{e^{-m_\pi r}}{r}
S_{12} \; .
\end{equation}
The transformation of this local potential
into momentum space yields for the $^3S_1$-$^3D_1$ amplitude
\begin{equation}
^\pi \tilde{V}^{110}_{02}(q_0,k) =
-\frac{g^2_\pi}{4\pi}
\frac{\sqrt{8}}{4\pi M^2 q_0 k}
[q^2_0 Q_2(z)-2q_0kQ_1(z)+k^2Q_0(z)]
\end{equation}
with $Q_L$ Legendre funct.\ of the 2.\
kind and $z\equiv (q^2_0+k^2+m^2_\pi)/(2q_0k)$.
The original $^3S_1$-$^3D_1$ transition potential
as it results from the relativistic one-pion-exchange
Feynman amplitude
is
\begin{eqnarray}
^\pi V^{110}_{02}(q_0,k) & = &
-\frac{g^2_\pi}{4\pi}
\frac{\sqrt{8}}{4\pi M^2 q_0 k}
[(E_{q_0}-M)(E_k+M) Q_2(z)-2q_0kQ_1(z) \nonumber \\
  &  &
+(E_{q_0}+M)(E_k-M)Q_0(z)]
\end{eqnarray}
with $E_{q_0}\equiv\sqrt{M^2+q^2_0}$
and $E_k\equiv\sqrt{M^2+k^2}$.
Expanding these roots in terms of $q^2_0/M$ and
$k^2/M$ yields
\begin{eqnarray}
^\pi V^{110}_{02}(q_0,k) & \approx &
-\frac{g^2_\pi}{4\pi}
\frac{\sqrt{8}}{4\pi M^2 q_0 k}
\left[\left(q^2_0 +\frac{q_0^2k^2}{4M^2}-\frac{q_0^4}{4M^2}\cdots\right)
Q_2(z)-2q_0kQ_1(z)  \right.
 \nonumber \\
   &  &
 \left.
+\left(k^2 + \frac{q_0^2k^2}{4M^2}-\frac{k^4}{4M^2} \cdots \right) Q_0(z)
\right] \; .
\end{eqnarray}
Keeping terms up to
momentum squared, leads to the
local approximation Eq.~(3).
The largest term in the next order
is $-k^2/(4M^2)Q_0(z)$ which damps the tensor
potential off-shell.

The thin short-dashed curve in Fig.~2 shows what is obtained when
in the Bonn-B potential the pion tensor force is made local
and the dotted curve results when both $\pi$ and $\rho$
are local.
The latter curve is very close the Paris curve (thick long-dashed line),
which explains the origin for the differences
between the Paris and Bonn potentials.
Another source for nonlocality in the Bonn potential
is the factor $M/\sqrt{E_{q_0}E_k}$ which is applied to
the Feynman amplitude to define the potential.
The nonlocality created by this factor has been investigated in
great detail by Gl\"ockle and Witala [7].
This factor is included in the solid, short-dashed, and
dotted curves
in Fig.~2.
When this factor is omitted in the latter case, the
widely-spaced dots are
obtained, which demonstrates an additional source of
nonlocality in the Bonn potential.

This work was supported in part by NSF-Grant PHY-9211607.

\begin{center}
REFERENCES
\end{center}
\small
\begin{tabular}{ll}
1. & F. Sammarruca, D. P. Xu, and R. Machleidt, Phys. Rev.
 C {\bf 46}, 1636 (1992).
\\
2. & J. L. Friar {\it et al.}, Phys. Lett. {\bf B311}, 4 (1993).
\\
3. & R. Machleidt, Adv. Nucl. Phys. {\bf 19}, 189 (1989).
\\
4. & D. Y. Song and R. Machleidt, unpublished.
\\
5. & M. Lacombe {\it et al.}, Phys. Rev. C {\bf 21}, 861 (1980).
\\
6. &  V. G. J. Stoks {\it et al.},
Phys. Rev. C {\bf 48}, 792 (1993).
\\
7. & W. Gl\"ockle and H. Kamada, {\it Nonlocalities in NN forces defined
through the} \\ & {\it Blankenbecler-Sugar equation},
preprint, University of Bochum (1994).
\end{tabular}

\newpage
\normalsize

\begin{center}
\bf Figure Captions
\end{center}

{\bf Figure 1.} The $\epsilon_1$ mixing parameter below 350 MeV
as predicted by the Bonn-B~[3] (solid line) and the Paris~[5]
(dashed) potentials. The solid dots represent the Nijmegen multi-energy
NN phase shift analysis [6].

\vspace*{.5cm}

{\bf Figure 2.}
Half off-shell $^3S_1$-$^3D_1$ transition potential
of the Bonn-B [3] (solid line) and the Paris [5] (thick dashed line)
potentials.
The thin short-dashed and dotted lines are
obtained when in the Bonn-B potential
the local approximation
is used for the $\pi$ and for both $\pi$ and $\rho$,
respectively.
When in the latter case also the $M/\sqrt{E_{q_0}E_k}$
factor is left out,
the widely-spaced dots are obtained.
The solid dot is the on-shell point ($k=q_0$)
at which all curves are the same.


\vspace*{3cm}

{\it The figures
are available upon request from
\begin{center}
{\sc machleid@tamaluit.phys.uidaho.edu}
\end{center}
Please include your FAX number or mailing address with your request.}

\end{document}